\documentclass[twocolumn,aps]{revtex4}

\begin{document}
\title{Stiffness and energy losses in cylindrically symmetric
superconductor levitating systems}
\author{Carles Navau\dag\ddag and Alvaro Sanchez\dag\footnote[3]{To
whom correspondence should be addressed (alvar.sanchez@uab.es)}}
\address{\dag\ Grup d'Electromagnetisme, Departament de
F\'{\i}sica, Universitat Aut\`onoma de Barcelona, 08192 Bellaterra,
Barcelona, Catalonia, Spain}
\address{\ddag\ Escola Universit\`aria Salesiana de Sarri\`a, Rafael
Batlle 7,
08017 Barcelona, Catalonia, Spain.}

\begin{abstract}
Stiffness and hysteretic energy losses are calculated for a
magnetically levitating system composed of a type-II superconductor
and a permanent magnet when a small vibration is produced in the
system. We consider a cylindrically symmetric configuration with only
vertical movements and calculate the
current profiles under the assumption of the critical state model.
The calculations, based on magnetic energy minimization, take
into account the demagnetization fields inside the superconductor and
the actual shape of the applied field. The dependence of stiffness
and hysteretic energy losses upon the different important parameters
of
the system such as the superconductor aspect ratio, the relative size
of
the superconductor-permanent magnet, and the critical current of the
superconductor are all systematically studied. Finally, in view
of the results, we provide some trends on how a system such as the
one studied here could be designed in order to optimize both the
stiffness and the hysteretic losses.
\end{abstract}



\maketitle

\section{Introduction}
Superconducting levitation results from the interaction between an
applied field and the currents in a
superconductor\cite{levinphys}. Macroscopically, a type-II
superconductor with strong pinning can be studied using the critical
state model\cite{bean}. Within this framework, currents with constant
density are
induced in the superconductor due to a variation of the applied field
it feels, either for a change in the value of a uniform field or
for a displacement of the superconductor in a non-uniform field. A
relative movement between a superconductor and a permanent magnet
will thus result in a magnetic force between the induced currents in
the superconductor and the field of the magnet. If this force
compensates the weight of the levitated object and the conditions for
stability hold, the object will stably and passively
levitate\cite{moonbook}.

Earnshaw proved that only diamagnetic materials can passively
and stably levitate\cite{earnshaw}. Although type-I superconductors
can levitate (actually, they can produce large levitation forces
because of their strong diamagnetism), the use of type-II
superconductors has become more useful because of their rigidity.
The rigidity arises from the pinning of the flux lines and allows a
continuous range of passive equilibrium positions\cite{rigidbrandt}.
Yet, the use of the type-II superconductors results in an
energetic loss due mainly to the magnetic hysteresis\cite{coombs}.

A superconducting bearing is a system composed, in its simplest form,
of a superconductor and a permanent magnet. The rigidity and the
losses are important factors in the design of a bearing.  Actually,
complicated types of bearings have been designed in order to
improve rigidity and losses of the whole system. An excellent review
on superconducting bearings can be found in \cite{hullreview}.

Measurements of stiffness\cite{cansiz,riise} and
losses\cite{terentiev,yanghull} in levitating
superconductor-permanent magnet
bearings have been recently published. From the theoretical point of
view, the study
of stability and losses in high T$_c$-permanent
magnet bearings is a complicated task, mainly due to the trouble of
finding the current penetration inside the superconductor when
the applied field is not homogeneous. Several models have been
presented.
In \cite{leva} and \cite{prb} we proposed approximate analytical
expressions for stiffness and losses, assuming both a small
superconductor and approximate demagnetizing
factors. In Ref. \cite{riise} the stiffness of a bearing with a thin
film superconductor was calculated, considering the
exact form of the applied field and assuming a fully penetrated thin
superconductor. 
Based on the interaction between dipole images, Kordyuk
\cite{kordyuk} calculated the force and energy losses on a permanent
magnet over a flat ideally hard superconductor. Using the same model,
Hull and Cansiz\cite{cansiz} calculated the stiffness of such a
system.

In Refs. \cite{prb1serie,prb2serie} a model to calculate the current
penetration of currents inside a finite cylinder, taking into account
both the demagnetization factors and the radial components (and
inhomogeneity) of the applied field, was presented. The model, based
on the critical state and on the minimization of magnetic energy, was
used for calculating  the current penetration and levitation
forces in the case of a cylindrically symmetric bearing system
\cite{prb2serie}. In
this work, we will extend the application of the model to calculate,
not only the current penetration and the levitation force, but the
rigidity and the hysteretic energy losses resulting from the reaction
of the system to an eventual vibration. We will study the dependence
of these properties upon both intrinsic parameters of the
superconductor (its critical current) and geometrical parameters
(aspect ratio of the superconductor and relative size between
superconductor and permanent magnet).

This paper is structured as follows. In Section II we review the
calculation model used, focusing on its application to the present
case of vertical vibrations. In Section III we define the stiffness
and the hysteretic losses and expound some general trends of these
parameters. Results for the case of cylindrical symmetry are
presented
in Section IV. Section V is devoted to the discussion of the
results and their use to optimize a real system. Finally, the
conclusions can be found in Section VI.

\section{Modelization}
We consider a zero-field cooled type-II cylindrical superconductor
(SC) of radius $R$ and length $L$ located, initially, very far from a
coaxial cylindrical permanent magnet with uniform magnetization in
the direction of its axis (PM). The PM has radius $a$, length $b$,
and magnetization $M_{PM}$. A scheme of the system is shown in
Fig. \ref{f:sketch}. We shall use cylindrical coordinates with the
origin
located at
the top face of the PM. We consider that the PM is fixed and that the
SC,
initially above it, can be moved towards or away from it. We define
the distance between the PM and the SC, $d$, as the $z$-height of the
bottom face of the SC. The general process of descending the SC from
a very distant position to a minimum distance, $d_{\rm min}$, and
then ascending it to far from the PM, $d \longrightarrow \infty$, is
called the major loop.

We want to simulate the response of the SC after a small vibration of
the system, produced at some point of the major loop. A vibration is
regarded as a change in the actual direction of the movement and the
subsequent return to the original position. This movement is referred
to as a minor loop (see Fig. \ref{f:sketch}).
In this work, we only consider cylindrically symmetric situations.
That is, we only consider movements of the SC along the axial
direction, during both major and minor loops.

The response of a finite cylindrical SC in the critical state and
after a movement in the presence of a non-uniform (but cylindrically
symmetric) applied field has been studied in \cite{prb2serie} using a
numerical procedure described in \cite{prb1serie}. The model includes
the following assumptions, besides the cylindrical symmetry: a) no
equilibrium magnetization for the SC is considered; b) the PM is not
affected by the presence of the SC; and c) the superconductor is
assumed to be in the critical state. (The force creep due to thermal
demagnetization has been recently studied by Qin et al
\cite{quinbrandt}).
We calculated in \cite{prb2serie} the levitation force during a
major loop and studied the principal characteristics of that force
and its dependence on important parameters of the system.
We now briefly review the main steps in the calculation of the
desired magnitudes. After a variation of the position of the
superconductor, we calculate the currents induced in the
superconductor by imposing that these currents tend to minimize
the magnetic energy of the superconductor. When a variation of the
direction of movement is produced, currents in reverse direction
begin to enter the superconductor. In this latter case, we calculate
the
response of the SC using the typical procedure of the critical state
\cite{clemsanchez}
keeping the already set currents and calculating the penetration of
new reverse ones in the same way as for the initial stage but
considering the variation of the field. 

For a given current distribution inside the superconductor, the
magnetic force between the currents and the external applied field
in the case of
cylindrical symmetry has only a $z$-component given by 
\begin{equation}
\label{eq.force}
F_z = 2 \pi \mu_0 \int_V J_c(\rho,z) H_\rho^a(\rho,z) \rho \; {\rm
d}\rho,
\end{equation}
where $V=\pi R^2 L$ is the volume of the SC and $H_\rho^a$ is the
radial component of the applied field created by the PM.

Although in the described procedure we could implement a dependence
of the critical current with the internal field, in this work we use
a constant critical current $J_c$, for the sake of simplicity. In
that case, Eq. \ref{eq.force} becomes
\begin{equation}
F_z = 2\pi \mu_0 J_c \int_\Omega H_r^a(\rho,z) \rho {\rm d}\rho,
\label{eq.force2}
\end{equation}
being $\Omega$ the current penetrated region. $\Omega$ depends not
only on the value of $J_c$ but on the geometry of the sample, because
of the demagnetizing fields. If the superconductor becomes fully
penetrated by currents, $\Omega=V$ and the vertical force is
proportional to $J_c$.

Unless the contrary is explicitly written, we will consider the
following
parameters: $a=b=R=L=0.01{\rm m}$, $J_c=2.81\;10^7 {\rm A/m^2}$, and
$M_{PM}=7.95\;10^5 {\rm A/m}$. The geometrical values have been
chosen among the typical ones in levitation experiments. The value
for $M_{PM}$ is such that $\mu_0 M_{PM}=$1 T. $J_c$ is chosen to
equal $H_0/R$, being $H_0$ the magnetic field at the origin of
coordinates (the center of the top face of the PM). Using these
parameters, the field $H_0$ can be considered as a typical field of
penetration of the superconductor since when the superconductor is at
zero distance (so the applied field in the center of its lower face
is $H_0$) the superconductor is
considerably penetrated by currents (although not completely
penetrated because the sample is finite and $H_p=J_cR$ is the
penetration field for an infinite sample in a uniform applied field).
We will see below why this is an appropriate
parameter.
The use of non-normalized variables allows the easier comparison with
experimental values. However, an adequate normalization could reduce
the number of parameters of the system (see Ref. \cite{leva}). In
this work, we will use non-normalized variables for the calculations
but will discuss the results in relation to important
dimensionless parameters.

In Fig. \ref{f:major} we show calculated results illustrating the
general
shape
of the major loop of the
force-distance curve together with some minor loops at several
starting distances. In the inset, a detail of one minor loop is also
showed. Since the penetration of currents in the critical state is a
hysteretic phenomenon, so the force is, in both the major and the
minor loops. 
In Fig. \ref{f:current} we show calculated current penetration
profiles
corresponding to several points
of one minor loop (we
have chosen the loop in
the inset of Fig. \ref{f:sketch}; the plotted points are marked
there).
We observe
the hysteretic process of current penetration causing the hysteresis
in the force during a minor loop. 
As we shall see in detail in the next sections one can analyze the
stability and the hysteretic losses of the system from the shape of
these minor loops.

\section{Definitions and general considerations} 
\subsection{Restoring force and stiffness}
A useful mechanical system, including those with levitating
components, is often required to be stable in all directions.
Any variation of the position of the working point of a rigid
solid must result in a restoring force
that tend to return the system to its original place. Considering
small displacements from the working position, one could define the
stiffness as the first order spring constant of the restoring force.
Thus, we define a stiffness matrix whose elements are
\begin{equation}
\label{eq.def.stif}
\kappa_{\alpha \beta}=-[\nabla_\beta F_\alpha]_{\rm r_0}.
\end{equation}
The physical meaning of the $\alpha \beta$-element is the following:
it is the first order spring constant of the restoring
$\alpha$-component of the force, after a variation of the working
point ${\bf r_0}$ in the $\beta$-direction. The minus sing is set by
convenience, as discussed below.

In this work we consider a cylindrical system in which the
working position and the vibrations of the system will always
maintain the
cylindrical symmetry. The force after one of such variations will
have only axial component, which results in $\kappa_{\rho
z}=\kappa_{\theta z}=0$. In that case, thus, only the
component $\kappa_{zz}$ should be accounted for. From 
Eq. \ref{eq.def.stif}, we have that
\begin{equation}
\label{eq.stifzz}
\kappa_{zz}(z_1)=- \left[\frac{\partial F_z^{ml}}{\partial z}
\right]_{z\rightarrow z_1}.
\end{equation}
The upper index $ml$ indicates that the force that should be
considered is the one within the minor loop. This is because, in the
case of type-II superconductors, the
force is hysteretic, so the force after the start of a vibration is
the one inside a minor loop.

Although in this work we only calculate the $\kappa_{zz}$ component,
it is
interesting to discuss some general properties of the stiffness
matrix in the case of a cylindrically symmetric working point. Since
there is no angular dependence, any displacement of the
superconductor in the angular direction should
not modify the current distribution so the restoring force is zero.
Thus, $\kappa_{\rho \theta} = \kappa_{\theta \theta} =\kappa_{z
\theta} = 0$. Moreover, by symmetry, any displacement in the radial
direction will tend to give a restoring force with no-angular
component, resulting in $\kappa_{\theta \rho}=0$. In that latter case
there will appear both a restoring force in the radial direction (and
stiffness $\kappa_{\rho \rho}$) which is the lateral restoring force,
and also a vertical force (with stiffness $\kappa_{z \rho}$) which
will tend to modify the height of the working point. 

The stiffness coefficients can have different values for the same
working point depending on how that working point has been achieved.
This is due to the hysteresis in the behavior of the levitation force
in type-II superconductors. Although the stiffness is only calculated
from the force during a minor loop, that force depends on the
previous magnetic history of the superconductor. In the case
considered here of vertical movements, we expect to obtain different
value for the $\kappa_{zz}(z_1)$ for a given $z_1$ when considering a
minor loop produced at a height $z_1$ during the general descending
stage or when considering a minor loop produced in the same $z_1$ but
in the major ascending stage.

Finally, the minus sign in Eq. \ref{eq.def.stif} is set for
convenience, since in this case the condition for vertical stability
is $\kappa_{zz}>0$. If, at some equilibrium position, an increment in
the
vertical distance is produced ($\partial z>0$), the levitation force
should decrease ($\partial F_z<0$; the weight of the object would
tend to return it to the equilibrium distance). On the other hand, if
the levitated object is pushed down ($\partial z<0$), the levitation
force should increase ($\partial F_z>0$). In both cases
$\kappa_{zz}>0$.

\subsection{Hysteresis and energy losses}
The hysteretic energy losses can be defined as the work done by the
system during a complete minor loop. They can alternatively be
defined as the energy needed to force a levitating object to follow a
minor loop and return to the original position. The only difference
is the sign of the two expressions. To avoid this ambiguity, we
define the hysteretic energy losses as 
\begin{equation}
\label{eq.def.damp}
E=\left| \oint_{ml} {\bf F} {\rm d}{\bf r} \right|,
\end{equation}
where the subindex $ml$ indicates that the integration should be done
over the minor loop.

This general expression can be simplified when considering that the
cylindrical symmetry is preserved through the minor loop. In that
case, $F_\rho=F_\theta=0$. If we consider that the minor loop runs
between $z_1$ and $z_2$ (with $z_1<z_2$) and name $E_{zz}$ the
resulting hysteretic energy losses, Eq. \ref{eq.def.damp} is
simplified into 
\begin{equation}
\label{eq.dampzz}
E_{zz}(z_1)=\int_{z_1}^{z_2}|F_z^\uparrow - F_z^\downarrow| {\rm d}z,
\end{equation}
where $F_z^\uparrow$ and $F_z^\downarrow$ refer, respectively, to the
force during the ascending and descending movements of the
superconductor during the considered minor loop. 

The point $z_1$ at which $E_{zz}$ is calculated is the starting point
of the vibration only if produced during the major descending stage.
If the minor loop is
produced during the major ascending stage, then $z_1$ is the
reversing point of the vibration. The reason is that,
differently from the stiffness coefficient $\kappa_{zz}$, the energy
losses at a given point do not depend (in the critical state
framework) on the previous magnetic history of the superconductor. In
the case considered here of vertical movements, $E_{zz}(z_1)$ does
not depend on whether $z_1$ is the one in the general descending
stage or in the ascending one. This fact can be understood from
Eq. \ref{eq.def.damp}. Since in the calculation of $E_{zz}$ one must
consider the difference between forces during the minor loop, the
eventual contribution of the frozen currents to the force is
eliminated from the equation, resulting in the 
non-hysteretic behavior of $E_{zz}$.

\section{Results for $\kappa_{zz}$ and $E_{zz}$.}
In this section we present calculations of the vertical stability and
the hysteretic energy losses of a cylindrically levitated system. We
systematically study the dependence of stiffness $\kappa_{zz}$ and
losses $E_{zz}$ on different relevant parameters of the system such
as the critical current of the SC, the aspect ratio of the SC, and
the relative dimensions of the SC and the PM. We consider in all
cases minor loops started at a fixed distance ($z_1=0.005{\rm m}$)
and amplitude of the vibration ($\Delta z = 0.0025{\rm m}$).

Both coefficients $\kappa_{zz}$ and $E_{zz}$ have some general
features independently of the parameters of the system (except in
some extreme limits). 
During the initial descending, both $\kappa_{zz}$ and $E_{zz}$
increase as $z_1$ decreases owing to the increasing of the field
gradient and the larger penetration\cite{eucas01}. As stated above,
$\kappa_{zz}$ is hysteretic whereas $E_{zz}$ is not. 
Another general feature is that, in the returning
stage, the stiffness becomes negative for a large enough $z_1$'s.
This fact,
already pointed out in \cite{leva}, is produced because for large
enough $z_1$'s the variation of the applied field is small and the
currents hardly enter inside the sample. The minor loops have then to
follow the major one, resulting in a negative stiffness (positive
initial slope). Actually, depending of the shape of the PM, it may
also
exists a non stable region in the major descending loop\cite{leva}.

\subsection{$J_c$ dependence}
\label{s.jcdep}
In Fig. \ref{f:keVsJc} we show the calculated $\kappa_{zz}$
(Fig. \ref{f:keVsJc}-a) and $E_{zz}$
(Fig. \ref{f:keVsJc}-b) as a function of the critical current $J_c$,
for
three
values of the aspect ratio $L/R$ of the SC, corresponding
respectively to a thin film sample ($L/R=1/5$), a regular size sample
($L/R=1$), and a large sample ($L/R=5$). In these calculations we
have fixed the PM and the value of $R$ (maintaining so the value
$a/R$).
We observe that the stiffness increases with increasing $J_c$,
whereas $E_{zz}$ has a maximum and tends to zero in both limits
for low-$J_c$ and high-$J_c$.

Results for the stiffness can be understood from a simplified model
that considers a small PM and a very thin
superconductor\cite{ieee7,riise}. Within this model, $\kappa_{zz}$
has two contributions:
\begin{equation}
\label{eq.stif.simp}
\kappa_{zz}(z_1)=-\mu_0\pi R^2 L \left[ -\chi_0 \left(
\frac{\partial H_z}{\partial z}\right)^2 + M_z(z_1) \frac{\partial^2
H_z}{\partial z^2}\right].
\end{equation}
The first term in the Eq. \ref{eq.stif.simp} is independent of
$J_c$ and is related to the initial susceptibility of the SC,
$\chi_0=\frac{8R}{3\pi L}$ \cite{clemsanchez}. The second one depends
upon the
magnetization of the SC and, thus, upon $J_c$. Riise et al
\cite{riise} reported that for a typical experimental setup the
second term is much less important that the first one, so, for their
very thin film, the stiffness was independent of $J_c$.

In the present general case, there are also two terms determining the
stiffness. Consider a minor loop produced during the descending of
the superconductor. The force, just at the starting point of
the minor loop, $z_1$, is
\begin{equation}
\label{eq.for.st1}
F_z^{fr}(z_1)=-2\pi \mu_0 \mid J_c \mid \int_0^L \int_{b(z',z_1)}^R
H_\rho^a(\rho,z') \rho {\rm d}\rho {\rm d}z',
\end{equation}
where $b(z',z_1)$ describes the current penetration profile.
If at this point a minor loop starts, these already induced current
are kept frozen (the superscript $fr$ in Eq. \ref{eq.for.st1}
indicates this) and new reversal currents begin to enter. Then, the
force can be calculated as
\begin{equation}
\label{eq.for.st2}
F_z(d)=F_z^{fr}(z_1)+F_z^{new}(d),
\end{equation}
where $F_z^{new}$ is the force due to the new induced currents:
\begin{equation}
F_z^{new}(d)=+2\pi \mu_0 \mid J_c \mid 2 \int_0^L \int_{a(z',d)}^R
H_\rho^a(\rho,z') \rho {\rm d}\rho {\rm d}z',
\end{equation}
being $a(z',d)$ the current profile of the new induced currents.

The coefficient $\kappa_{zz}$ is calculated from
Eq. \ref{eq.for.st2} as
\begin{equation}
\label{eq.for.st4}
\kappa_{zz}(z_1) = - \left[ \frac{{\rm d}F_z^{fr}}{{\rm d}d} +
\frac{{\rm d}F_z^{new}}{{\rm d}d}{} \right]_{d\longrightarrow z_1}.
\end{equation}
Although the currents induced prior to the minor loop are kept
frozen, the first term in Eq. \ref{eq.for.st4} is different from
zero because the applied field involved in the integral changes when
varying $z$. This term corresponds to the $M_z(z_1)
\frac{\partial^2
H_z}{\partial z^2}$ term in Eq. \ref{eq.stif.simp}. It depends on
$J_c$.
On the other hand, the second term in Eq. \ref{eq.for.st4} depends
on the initial slope of the magnetization of the superconducting
sample. It is the equivalent to the non-hysteretic term in
Eq. \ref{eq.stif.simp}. Within the critical state model, this term
does
not depend on $J_c$.

Considering a minor loop started at a given working point, increasing
$J_c$ increases the magnetization of the SC
sample. The first term in Eq. \ref{eq.for.st4} increases whereas
the second one does not. The result is an increase of the stiffness,
as can be seen in Fig. \ref{f:keVsJc}-a. 

In the large $J_c$ limit the force is
almost non-hysteretic. Therefore, in this limit, the minor loop runs
over the major one and the stiffness can be approximately
calculated directly from the slope of the major loop at a given
distance.

The results for $E_{zz}$ can be understood by looking at
Fig. \ref{f:mlVsJc},
where we have plotted several minor loops corresponding to the cases
$L/R=0.2$ (left), $L/R=1$ (middle), and $L/R=5$ (right), for
different $J_c$ values. (All the discussion below holds for the three
figures, although the complete discussion is best seen in
Fig. \ref{f:mlVsJc}-a).
For the case that $J_c$ is small (actually, when it is small enough
to produce a full penetration of the sample for a small variation in
$d$) as soon as the minor loop is started, the sample is fully
penetrated and as soon as the minor loop is reversed, the sample is
again fully penetrated by currents circulating in the opposite
direction. The result is an almost symmetric (with respect to the
zero-force) behavior of the force during the minor loop. In that
case, the levitation force is roughly proportional to the critical
current (Eq. \ref{eq.force2}) and, from Eq. \ref{eq.dampzz}, we can
see
that the energy
losses are 
\begin{equation}
\label{eq.ezz.lowjc}
E_{zz} \simeq 4\pi \mu_0 \mid J_c \mid \int_{z_2}^{z_1} \mid g(h)
\mid {\rm d}h,
\end{equation}
where $g(h)$ is defined as 
\begin{equation}
g(h)=\int_0^R \int_h^{h+L} H_\rho^a (\rho,z) \rho {\rm d}\rho
{\rm d}z.
\end{equation}
For low $J_c$, thus, $E_{zz}$ tends to zero, proportionally to $J_c$.
As $J_c$ increases, the full penetration is obtained only after a
significant portion of the minor loop has been gone through. The
minor loop
becomes less and less symmetric, but at the same time, the value of
the
force increases. The resulting area of the minor loop increases for a
range of not very large $J_c$'s. When $J_c$ goes on increasing, no
full penetration of reverse currents is achieved. Actually, for large
$J_c$'s, the superconductor is hardly penetrated, either for direct
currents or for reverse ones, so the losses are being reduced even
though the value of the force increases. The resulting $E_{zz}$ is
thus tending to zero for large $J_c$.

In Fig. \ref{f:keVsJc}-b we observe that the maximum in the $E_{zz}$
coefficient is
displaced to lower $J_c$ when the SC is larger. This is
understood again in terms of the penetration. For longer
superconductors (Fig. \ref{f:mlVsJc}-b,c), the demagnetization fields
are
lower
and, as a consequence, the running distance necessary to fully
penetrate the SC is larger. The described crossover behavior of
producing less hysteresis because of less penetration is attained,
thus, at lower $J_c$'s. A more detailed description of the dependence
on $L/R$ will be given in Section \ref{s.jcdep}.

\subsection{$L/R$ dependence}
\label{s.lrdep}
In Fig. 6 we plot the dependence of $\kappa_{zz}$
(Fig. \ref{f:keVslr}-a) and
$E_{zz}$ (Fig. \ref{f:keVslr}-b) on $L/R$ for different values of the
critical
current. In order to study the dependence only on $L/R$ we have fixed
the PM, the value of the $R$ (maintaining so the relation $a/R$), and
also the starting point ($z_1=0.005{\rm m}$) and the amplitude
($\Delta z=0.0025{\rm m}$) of the minor loops.

We observe in Fig. \ref{f:keVslr}-a that the stiffness
significatively
depends on
$L/R$ for low-$J_c$ but hardly for high-$J_c$. The coefficient
$E_{zz}$ has a singular behavior. As seen in Fig. \ref{f:keVslr}-b,
for
low
critical current, $E_{zz}$ increases for low $L/R$, and saturates for
long $L/R$. On the other hand, for high $J_c$, it decreases all the
time (within the $L/R$ range shown), even for small $L/R$, saturating
for long samples. 

The argument exposed in the previous section explains also the
behavior
of $\kappa_{zz}$ in the following way. The term in the stiffness
depending on the magnetization is more important for high-$J_c$ and,
in this case, the magnetization slightly depends on $L/R$ since
almost no current penetrates the SC specially far from the PM. So,
increasing $L$ does not produces
significative changes (except when the SC is very thin). On the other
hand, when $J_c$ is low, the sample is significatively penetrated by
currents and the relation $L/R$ plays an important role.

The behavior of $E_{zz}$ can be more clearly understood by looking at
Fig. \ref{f:mlVslr}-a and Fig. \ref{f:mlVslr}-b, where we have
plotted the
minor loops for
different lengths and for a case of low critical current
(Fig. \ref{f:mlVslr}-a)
and a case of higher critical current (Fig. \ref{f:mlVslr}-b). When
the
current is
low the force during a minor loop is almost symmetric with respect to
zero-force value (Section \ref{s.jcdep}). Even in this low-current
limit,
when
the superconductor is long, there are regions free of current, so the
behavior departs from the symmetrical one. The departure, however, is
not very strong since the applied field in the regions where no fully
penetration is achieved is weak. In Fig. \ref{f:mlVslr}-a we observe
this
behavior
and see that $E_{zz}$ tends to increase as $L/R$ decreases. The
appearance of a maximum in the $E_{zz}$ versus $L/R$ curve for this
case has the same origin as discussed in the previous
section (a similar behavior of that shown in Fig. \ref{f:mlVsJc}
could be
observed
in this case, but as function of $L/R$; the difference is that now,
when L/R is large, the minor loops tend to saturate to a constant
shape instead of changing indefinitely). When the critical current is
high, only when $L/R$ is small the penetration is complete because
the demagnetization effects are important in that case. When
increasing $L/R$, the minor loops get less and less hysteretic until
their shape (and thus, their area) is saturated (see
Fig. \ref{f:mlVslr}-b).
Actually,
even in the case of high-$J_c$ for thin enough samples the area of
the minor loop should tend to zero, producing also a maximum in the
force. In the scale used in Fig. \ref{f:mlVslr}-b, this is not seen.

Since both $\kappa_{zz}$ and $E_{zz}$ saturate for long
superconductors, their value per unit volume decrease as $1/L$ for
long enough superconductors.

\subsection{$a/R$ dependence}
We study in this section the dependence of $\kappa_{zz}$ and $E_{zz}$
upon the ratio between the radii of the PM and the SC, $a/R$. We
have maintained the value of $R$ in order to not change the relation
$L/R$. The value of $b$ is also maintained and so is $M_{PM}$. We
note, however, that when changing $a$, the applied field will change.
In
Fig. \ref{f:keVsar}-a and Fig. \ref{f:keVsar}-b we show,
respectively, the
calculated $\kappa_{zz}$
and $E_{zz}$ for three cases of $J_c$ corresponding to 0.1, 1, and 10
times the relation $H_0/R$.

In Fig. \ref{f:keVsar}-a we observe that there is a region of maximum
stability
(maximum value of stiffness)
when the superconductor and the permanent magnet have similar radius
(in Ref. \cite{prb2serie} it was shown that also the magnitude of the
force is larger
when $a\simeq R$). $E_{zz}$ shows a
maximum at $a\simeq R$ and tends to zero for both $a<<R$ and $a>>R$
limits. 

The results can be explained as follows. Given a magnetization
$M_{PM}$ of the permanent magnet, when the magnet has a radius
$a\longrightarrow 0$ the value of the applied field tends to zero and
also does the levitation force. Both stiffness and energy losses
should be zero in that limit. On the other hand, when the PM is much
larger than the SC, in the region occupied by the superconductor the
field is approximately homogeneous, resulting also in an almost null
levitation force and, as a consequence, a very low stiffness and
losses.
In the intermediate regions, there should be, at least, one maximum.
The fact that the maximum corresponds to $a \simeq R$ is related to
the fact that, at a given distance from the PM, the applied field is
most inhomogeneous at radial distances in the range of $a$.
Therefore, SC's of radius $R\simeq a$ will feel larger variations in
the applied field, resulting in a larger penetration of currents,
which
means a larger rigidity and, at the same time, larger losses, because
of the larger hysteresis.

\section{Discussion}
The levitation force of type-II superconductors is the result of the
interaction between a distribution of induced critical
currents and the external applied field. Stiffness and losses
can be interpreted in terms of the current
penetration profiles. In a qualitative way, the key parameter that
accounts for the behavior of a bearing system is the relation between
the field of penetration of the SC and the applied field of the PM.
Other parameters are related with this. The penetration field for the
superconductor depends on the critical current as well as on its
dimensions and aspect ratio. (In this work we have only worked 
with cylinders, but in general, the penetration field also depends on
the geometry of the superconducting sample). Actually, the larger the
critical current is, the larger the penetration field is
(proportionally) and the smaller $L/R$ is, the larger that
penetration field is (not in a proportional way \cite{forkl}). The
applied field felt by the superconductor depends on the
magnetization of the PM (proportionally, if considered uniform and
constant), on the dimensions of the PM and on the distance between
them. 

The relation between these two fields determines the current
penetration profiles in the following way: if the applied field felt
by the SC is much smaller than its penetration field, current will
hardly penetrate the SC, which will yield in almost non-hysteretic
behavior and large forces.  This condition can be achieved by several
reasons (or by a combination of them), namely, high $J_c$, large
$L/R$, low $M_{PM}$, small $a$ (in relation to $R$), and/or small
$b$. 
On the other hand, if the applied field felt by the SC is much larger
that its penetration field, currents will penetrate a considerable
amount of SC material after a small field variation (after a small
movement of the SC). This will result in a highly hysteretic
behavior. This condition will be fulfilled for low $J_c$, small
$L/R$, high $M_{PM}$, and/or $a\simeq R$. 

For the quantitative description of the stiffness and losses one
should also take into account the actual value of the force. For
example,
even when there is large hysteresis, the area of a minor loop
can be small if the applied field (and the force) is small
(Fig. \ref{f:mlVsJc}-a). Also, large hysteresis (almost symmetrical
minor
loops) does not
necessarily means a large stiffness. We have seen that large $J_c$'s
increases the initial slope of a minor loop (in absolute value)
although the hysteresis is reduced (Fig. \ref{f:mlVsJc} and
Fig. \ref{f:mlVslr}).

\section{Conclusions}
Using a model based on the critical-state approximation, we
have been able to calculate the response of a levitating type-II
superconductor in the presence on the non-uniform applied field
created by a permanent magnet. We have focused our analysis on
vertical vibrations that can occur at a given working point. 

We have seen that 
when the radius of the PM and SC are similar, the stability of the
system is maximized, although the losses are also the largest. 
As to the length of the superconductor, a very long sample
can have a significant amount of material that contribute
neither to the force, nor to the stiffness, nor to the hysteretic
energy losses,
resulting therefore in a useless waste of SC material. By increasing
the critical
current of the superconductor, the stability of the system is
increased, whereas the losses could be reduced, if this critical
current is high enough.

Optimization of a cylindrical bearing system should take into account
all these results. Depending on the needs and the restrictions for a
particular design, one should choose a permanent magnet and a
superconductor that fit the needs. In particular, very high
critical current could produce large stability with low losses.
However if a thin film SC is used, even with such high critical
current the sample could be easily fully penetrated and the losses
would be
important. If large levitation forces are
needed, a long
superconductor with high critical current would be the optimum
solution for the
reduction of losses, without considerably reducing the stability.

In its current form, the
present model is unable to describe some important phenomena of
levitation, such lateral displacements (and lateral stiffness and
hysteretic losses), movements at high frequencies (such that the
critical state does not have time to be established
inside the SC), force creep, or even finite geometries other than
cylindrical. However, the same methodology could be eventually
applied to these cases. If by some means, a procedure to obtain the
current
profiles inside the superconductor is found, the procedure described
would be valid for a phenomenological description of
the levitation force. In particular, the relation between a typical
penetration field of the superconductor and a typical applied field
is expected to be the key parameter to explain the 
experimental data, even in situations different from the cylindrical
symmetry described here.

\section*{Acknowledgements}
We thank MCyT project BFM2000-0001 and CIRIT project
1999SGR00340 for financial support.


\newpage
\begin{figure}
\caption{\label{f:sketch}Geometry and dimensions of the considered
superconductor-permanent magnet system (left).
Sketch of a minor loop running between $z_1$ and $z_2$ (right).}
\end{figure}

\begin{figure}
\caption{\label{f:major}Vertical levitation force for a complete
major
loop and for
several minor loops of amplitude $\Delta z$=0.005m produced every
0.005m. Other parameters of the calculated case are in the text. In
the inset there is a zoom of one of the minor loops. The current
penetration profiles in the marked points are shown in
Fig. \ref{f:current}.}
\end{figure}

\begin{figure}
\caption{\label{f:current}Current penetration profiles for a semi
plane of constant
angle (the $z$-axis is at left in every figure) at several
distances during a minor loop. The distances are the ones marked in
Fig. \ref{f:major}. Currents induced during the descend of the minor
loop
(increase of the applied field) are drawn in dark gray. In light gray
there are the reverse currents induced during the ascent of the
superconductor.}
\end{figure}

\begin{figure}
\caption{\label{f:keVsJc}(a) Vertical stiffness $\kappa_{zz}$ and (b)
hysteretic
energy losses $E_{zz}$ as a function of the critical current of the
superconductor. The PM is fixed and the aspect ratio of the
superconductor is changed by changing the value of $L$ such that
$L/R$=0.2 (dotted line), $L/R$=1 (solid line), and $L/R$=5 (dashed
line). Other parameters are written in the text.}
\end{figure}

\begin{figure}
\caption{\label{f:mlVsJc}Force versus distance minor loops started at
$z_1$=0.005m
and of amplitude $\Delta z$=0.0025m for the cases (a) $L/R=0.2$, (b)
$L/R=1$, and (c) $L/R=5$. The values of $J_c$ have been chosen to be
equally spaced in logarithm scale and they range (from lower to
higher
maximum force ---thinner to thicker line) (a) from
$J_c=1.000\;10^6{\rm A/m^2}$ to $J_c=5.623\;10^7{\rm A/m^2}$;
(b) and (c) from $J_c=1.000\;10^7{\rm A/m^2}$ to $J_c=5.623\;10^8{\rm
A/m^2}$. For comparison, the loop corresponding to $J_c=
1.000\;10^7{\rm A/m^2}$ is showed as a dotted line.}
\end{figure}

\begin{figure}
\caption{\label{f:keVslr}(a) Vertical stiffness $\kappa_{zz}$ and (b)
hysteretic
energy losses $E_{zz}$ as a function of the aspect ratio of the
superconductor, for different values of the critical current:
$J_cR/H_0$=0.1 (solid line),
$J_cR/H_0$=1 (dot-dashed line). Other parameters are written in the
text.}
\end{figure}

\begin{figure}
\caption{\label{f:mlVslr}Force versus distance minor loops started at
$z_1$=0.005m
and of amplitude $\Delta z$=0.0025m for the cases (a)$J_cR/H_0$=0.1
and (b) $J_cR/H_0$=1. The value of $L/R$ is (from lower to higher
maximum force ---thinner to thicker line) $L/R=0.1$, $0.5$, $1.0$,
$1.5$, and $2.0$, respectively.}
\end{figure}

\begin{figure}
\caption{\label{f:keVsar}(a) Vertical stiffness $\kappa_{zz}$ and (b)
hysteretic
energy losses $E_{zz}$ as a function of the relative radius of the PM
and the superconductor $a/R$, for different values of the critical
current: $J_cR/H_0$=0.1 in dotted line and $J_cR/H_0$=1 in solid
line. Other parameters are written in the text.}
\end{figure}

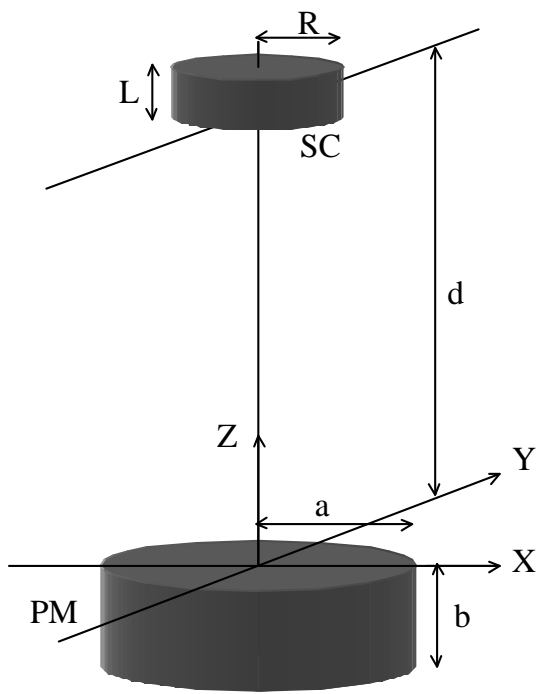
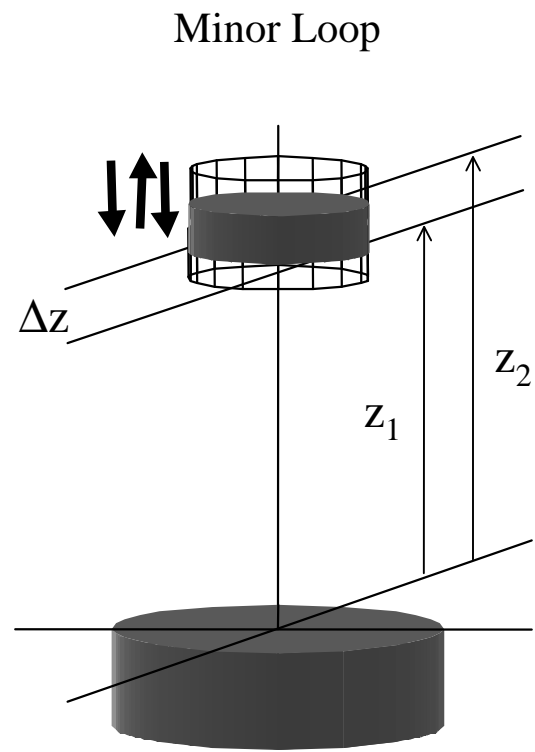
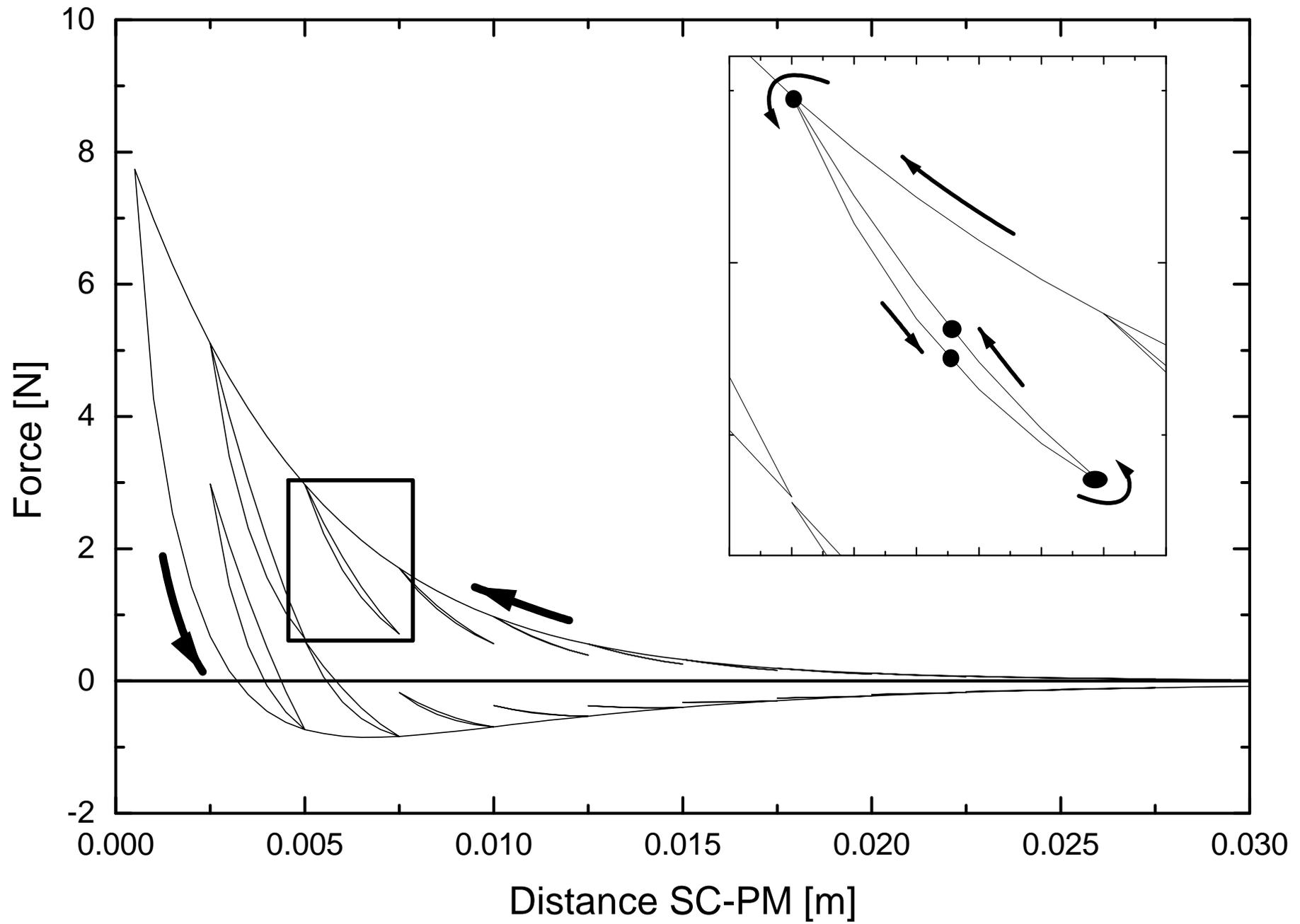
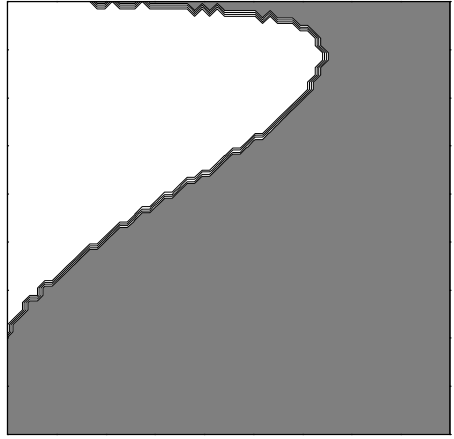
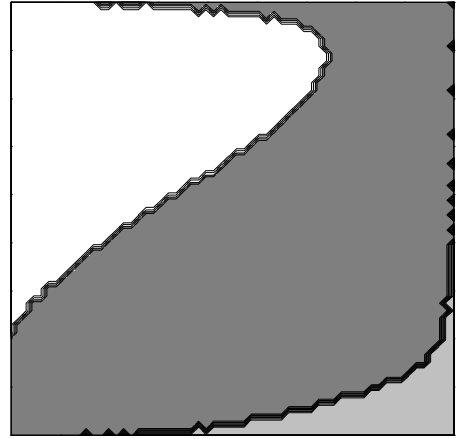
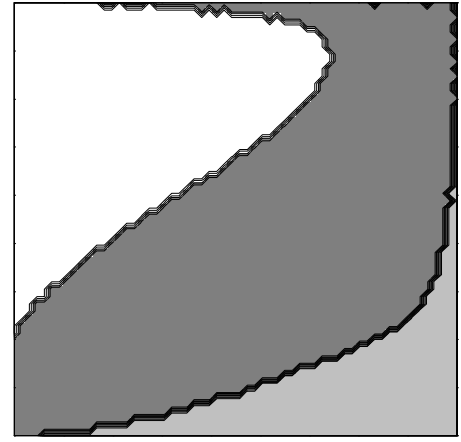
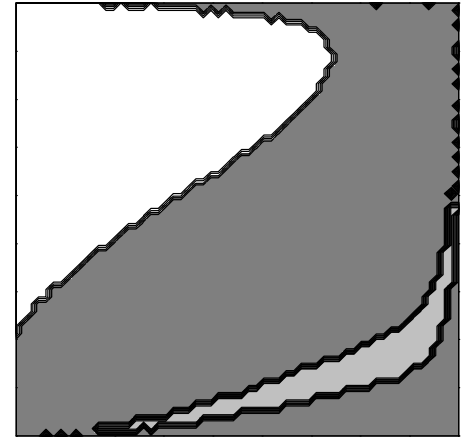
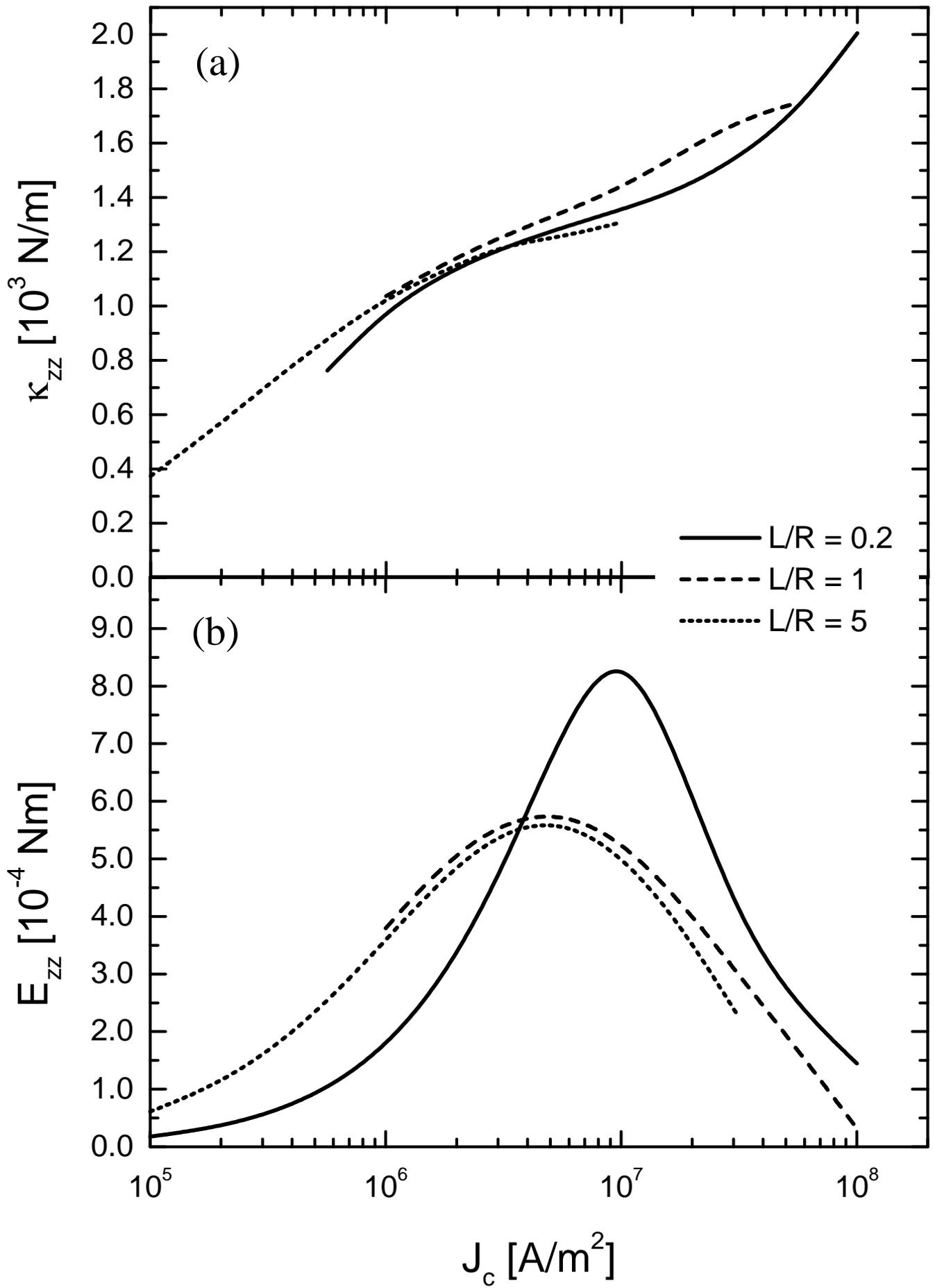
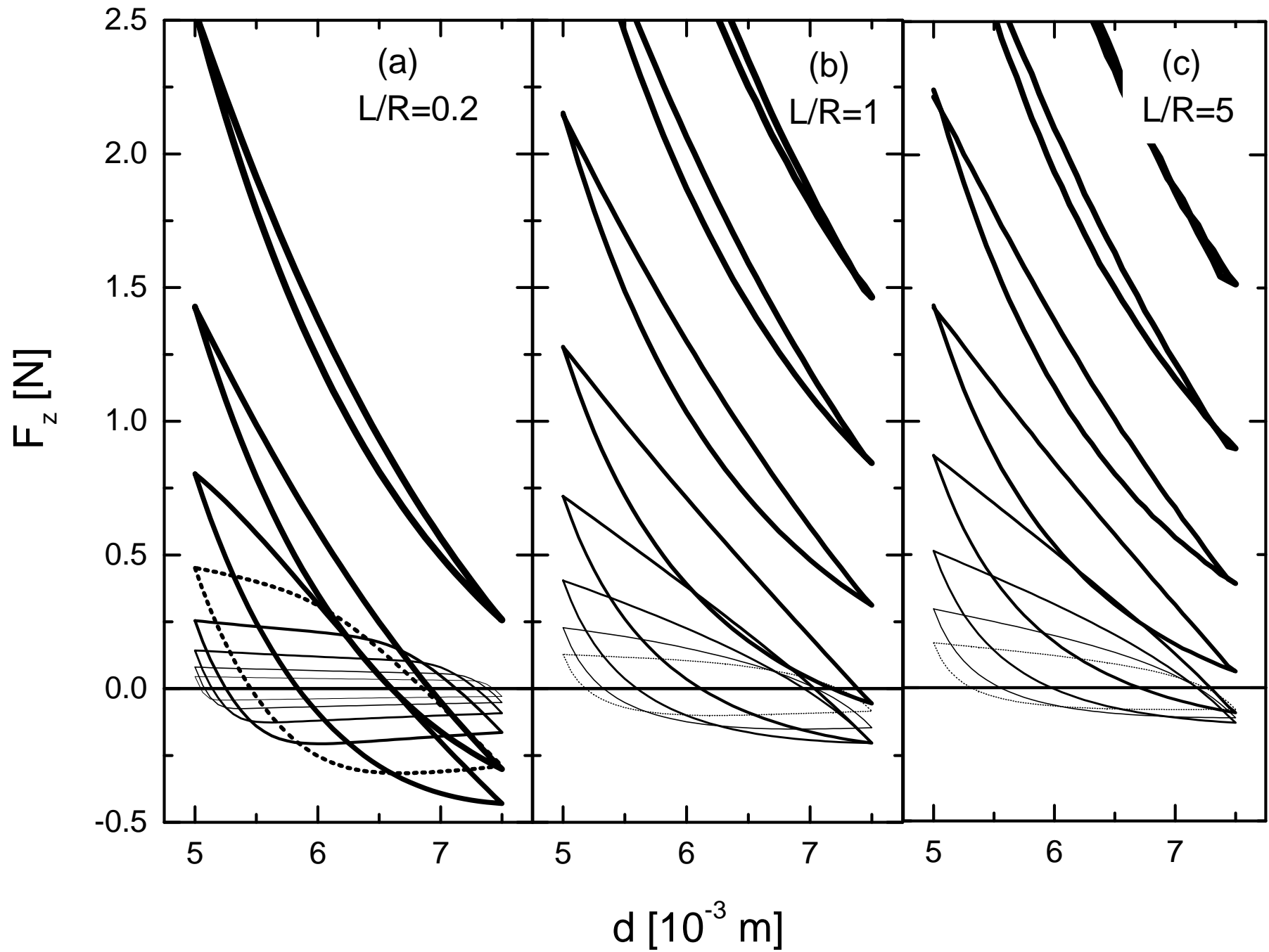
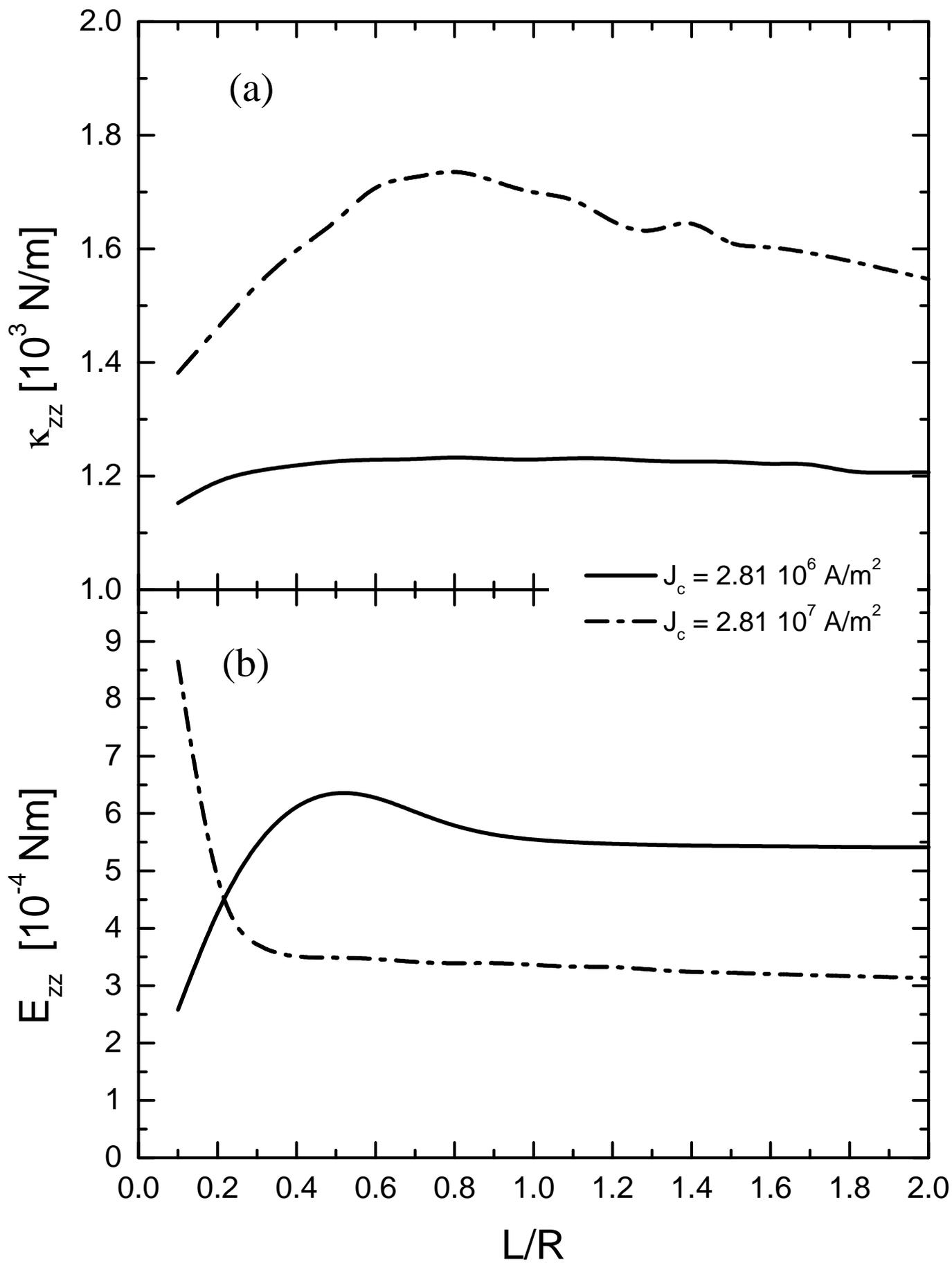
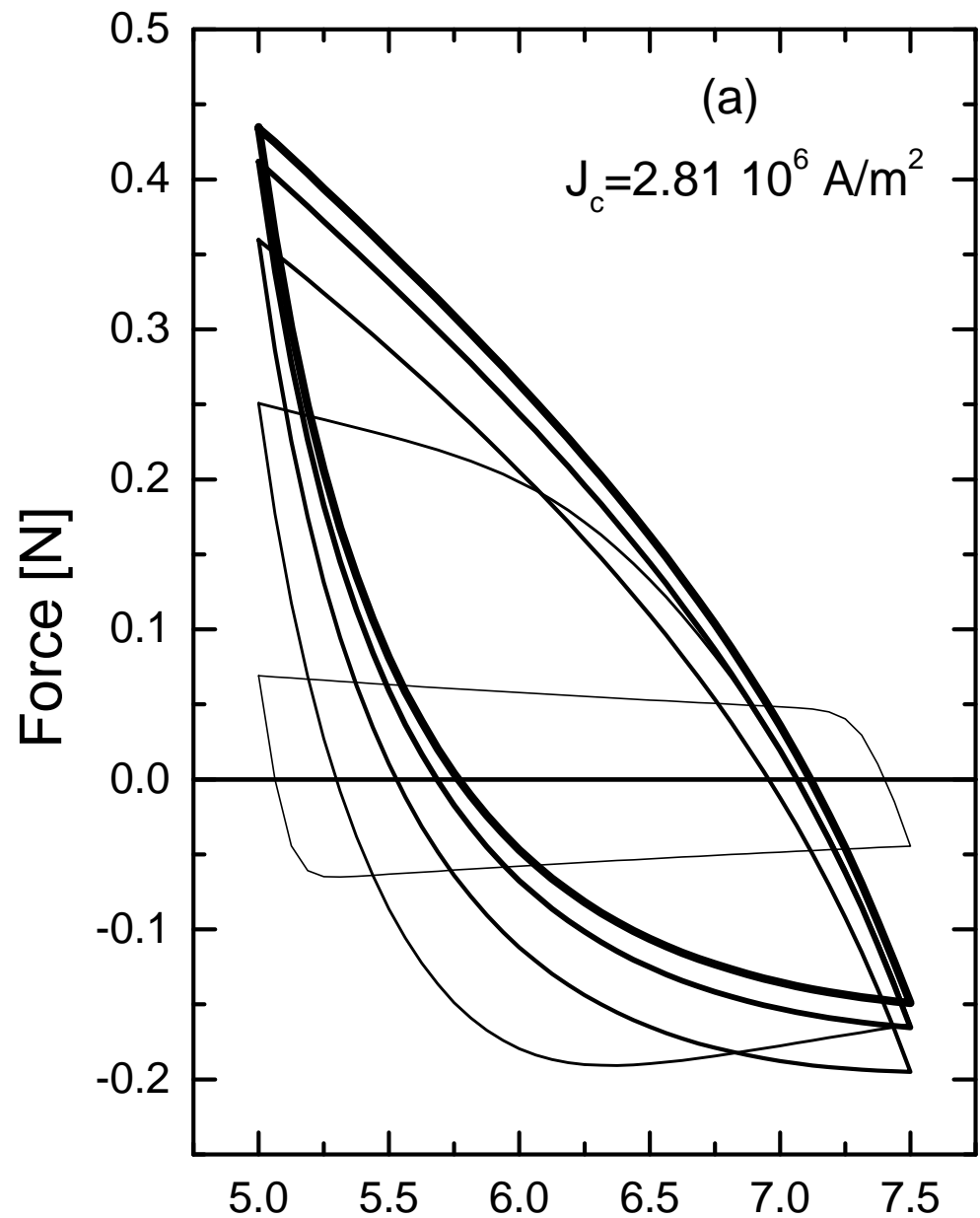
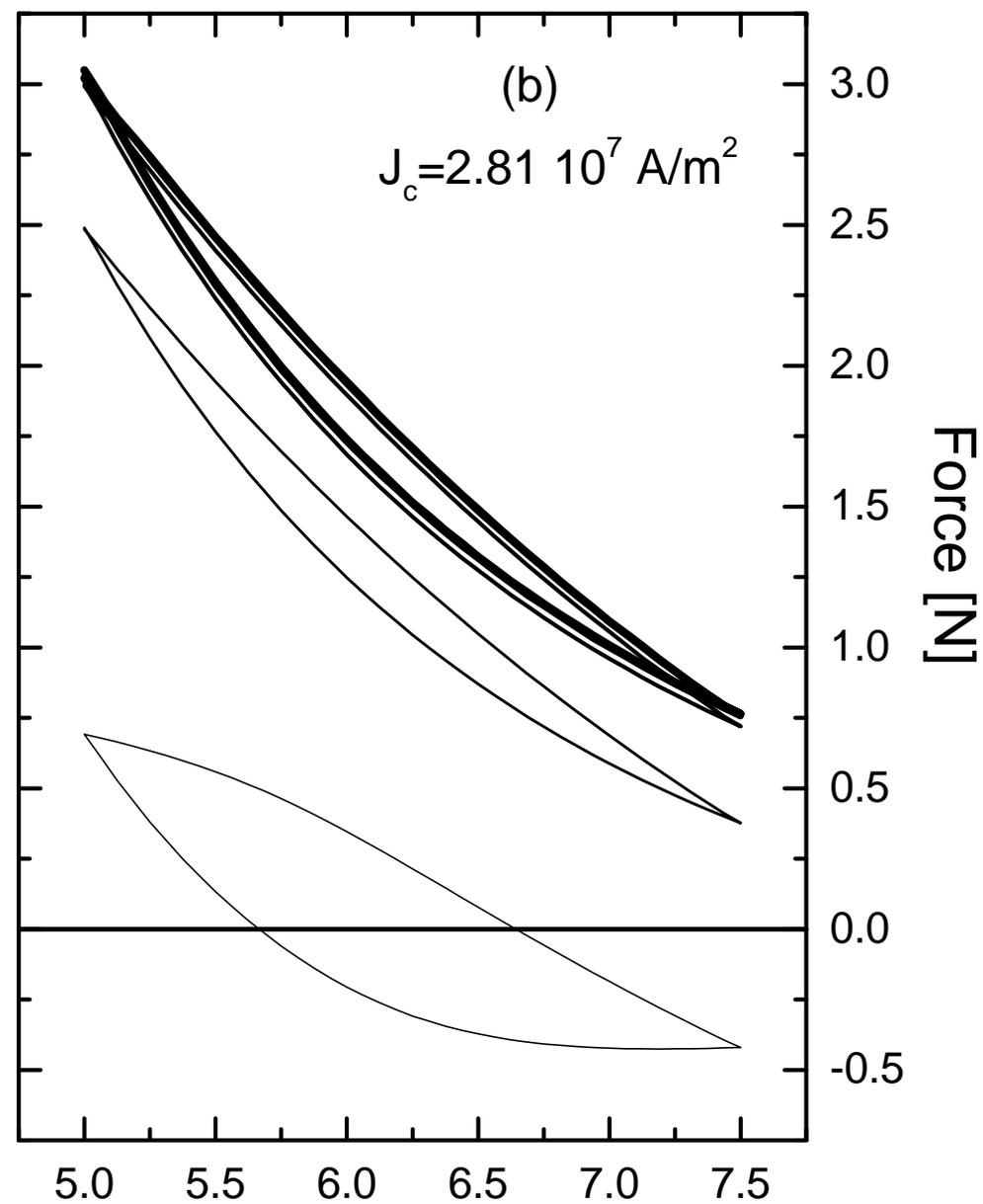
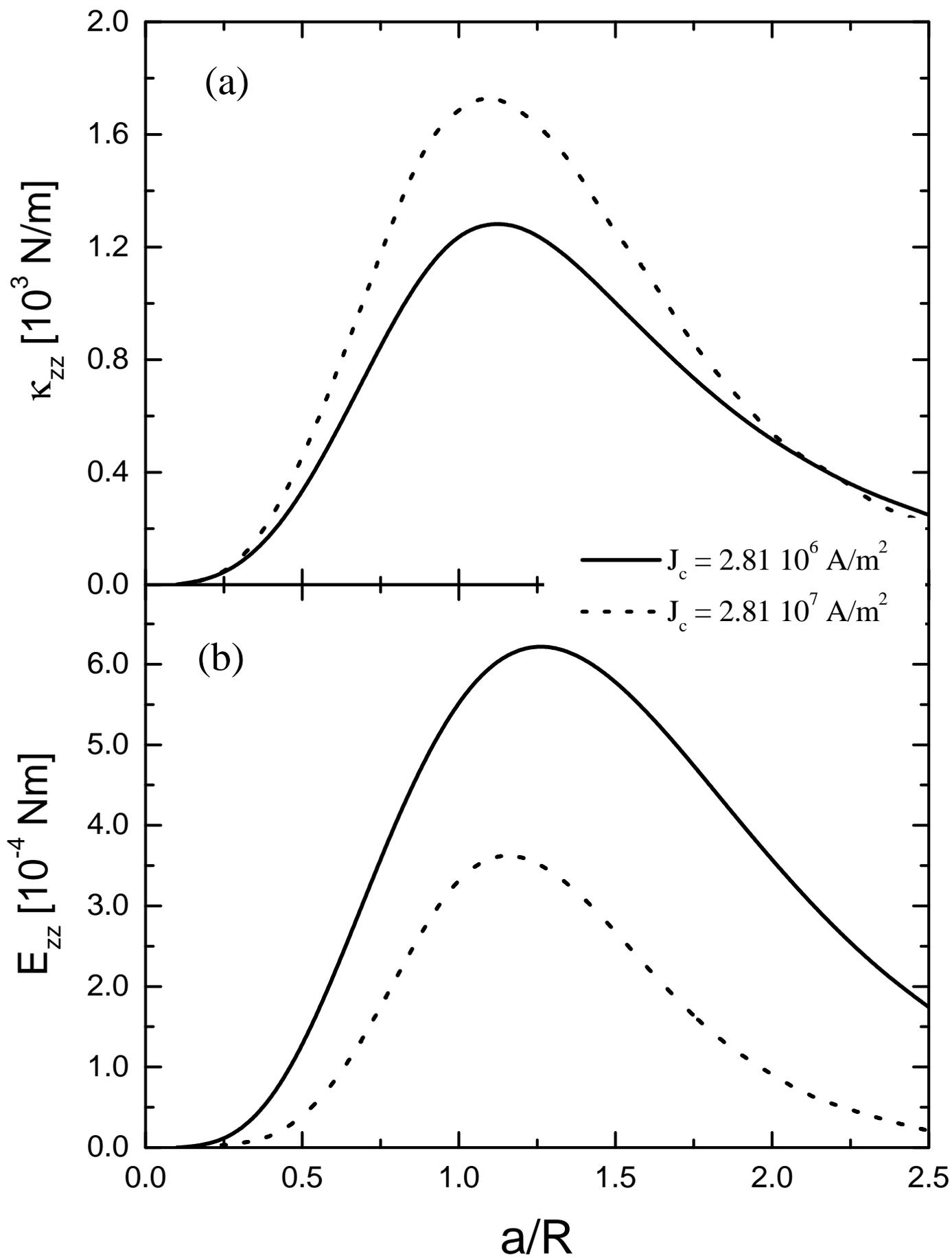
\end{document}